\documentclass[12pt]{article}
\usepackage{bbm}
\usepackage{bbding}
\usepackage{amsmath}
\usepackage{amssymb}
\usepackage{amsfonts}
\usepackage{mathrsfs}
\usepackage{mathrsfs, amssymb}
\usepackage{booktabs}
\usepackage[perpage,symbol*]{footmisc}
\usepackage{graphicx}
\usepackage{multirow}
\usepackage[level]{datetime}
\usepackage[a4paper]{geometry}

\newcounter{theorem}
\newtheorem{theorem}{Theorem}[section]
\newcounter{lemma}

\newcounter{remark}

\newcounter{example}

\newcounter{definition}

\newcounter{corollary}

\newcounter{proposition}

\newcounter{assumption}

\newcounter{condition}

\newcounter{algorithm}

\makeatletter
\newcommand*{\indep}{%
  \mathbin{%
    \mathpalette{\@indep}{}%
  }%
}
\newcommand*{\nindep}{%
  \mathbin{
    \mathpalette{\@indep}{\not}
  }%
}
\newcommand*{\@indep}[2]{%
  \sbox0{$#1\perp\m@th$}
  \sbox2{$#1=$}
  \sbox4{$#1\vcenter{}$}
  \rlap{\copy0}
  \dimen@=\dimexpr\ht2-\ht4-.2pt\relax
  \kern\dimen@
  {#2}%
  \kern\dimen@
  \copy0 
}
\makeatother

\def\vbeta{\mbox{\boldmath$\beta$}}
\def\vdelta{\mbox{\boldmath$\delta$}}

\def\vtheta{\mbox{\boldmath$\theta$}}

\def\vmu{\mbox{\boldmath$\mu$}}

\def\vSigma{\mbox{\boldmath$\Sigma$}}

\def\a{\mbox{\boldmath$a$}}
\def\A{\mbox{\boldmath$A$}}
\def\B{\mbox{\boldmath$B$}}
\def\D{\mbox{\boldmath$D$}}
\def\e{\mbox{\boldmath$e$}}
\def\G{\mbox{\boldmath$G$}}

\def\b{\mbox{\boldmath$b$}}

\def\u{\mbox{\boldmath$u$}}
\def\U{\mbox{\boldmath$U$}}
\def\V{\mbox{\boldmath$V$}}

\def\I{\mbox{\boldmath$I$}}
\def\0{\mbox{\boldmath$0$}}
\def\H_N{\mbox{\boldmath$H_N$}}

\def\tr{\mbox{tr}}

\def\drow{\stackrel{d}{\longrightarrow}}

\newcommand{\bmsection}[1]{%
	\section{\textbf{#1}}%
}

\begin{document}
\baselineskip 18pt
\begin{center}
{\large\bf Optimal subsampling for functional composite quantile regression in  massive data}
\end{center}

\begin{center}

{ \textbf{\small{ Jingxiang Pan}}}\\
{ \small{School of Mathematics and Statistics, Changchun University of Technology, China}}\\
{ \small{\texttt{\textsc{pjx844227330@gmail.com}}}}

{ \textbf{\small{ Xiaohui Yuan}}$^*$}\\
{ \small{School of Mathematics and Statistics, Changchun University of Technology, China}}\\
{ \small{\texttt{\textsc{yuanxh@ccut.edu.cn}}}}

{ \textbf{\small{ Danyang Wang}}}\\
{ \small{School of Mathematics and Statistics, Changchun University of Technology, China}}\\
{ \small{\texttt{\textsc{wdy\_ynl@163.com}}}}

\end{center}

\begin{center}
This version: \usdate\today
\end{center}

\footnotetext{$^*$Corresponding author, $^\dag$ equal authors contribution.}

\begin{abstract}
{As computer resources become increasingly limited, traditional statistical methods face challenges in analyzing massive data, especially in functional data analysis. To address this issue, subsampling offers a viable solution by significantly reducing computational requirements. This paper introduces a subsampling technique for composite quantile regression, designed for efficient application within the functional linear model on large datasets. We establish the asymptotic distribution of the subsampling estimator and introduce an optimal subsampling method based on the functional L-optimality criterion. Results from simulation studies and the real data analysis consistently demonstrate the superiority of the L-optimality criterion-based optimal subsampling method over the uniform subsampling approach.
}
\end{abstract}

\section {Introduction} \label{sec1} \setcounter {equation}{0}
\def\theequation{\thesection.\arabic{equation}}

Functional data, stored in the form of functional curves, have appeared in many fields such as climate, finance, economy, and biology. In functional data analysis, extensive research has been conducted on the functional linear model to explore the relationship between functional variables and a scalar response, such as Yao et al.  (2005) , Peter \& Joel (2007) , Hilgert et al.  (2013) , Aneiros et al. (2022) and so on. Since most functional data have features such as heteroscedasticity, large fluctuation amplitude, and inconvenient description of tail features, the composite quantile regression model and quantile regression model were introduced into the analysis of functional data, such as Kato (2012) , Yu et al. (2016) , Ma et al. (2019), Ping et al. (2021), Sang \& Cao (2022), Li et al. (2022) and so on.

As technology advances the cost of collecting and storing data becomes lower, access to massive amounts of data becomes easier and easier. The emergence of massive functional data rendering existing methods for processing functional data infeasible. Subsampling algorithms provide an effective means of reducing the computational burden when analyzing large amounts of data. This approach involves taking a small subsample from the entire dataset for estimation. Subsampling algorithms have been extensively studied and proven effective in the academic literature. Ma et al. (2015) investigated the bias and variance of least squares regression estimators based on subsampled samples.  Wang proposed a subsampling algorithm aimed at efficiently approximating maximum likelihood estimation within a logistic regression framework. The study delves into the estimates obtained by the subsampling algorithm and successfully verifies the consistency and asymptotic normal distribution properties of these estimates. In addition, Wang systematically derives the optimal subsampling probability, which is determined to minimize the mean square of the estimation error in an asymptotic sense.
Ai et al. (2021) explored optimal subsampling methods based on the A-optimality criterion in generalized linear models with the aim of efficiently approximating maximum likelihood estimates for large-scale data sets. Yu et al. (2022) proposed the Poisson subsampling methods for quasi-likelihood estimation, in reducing computational burden and maintaining estimation efficiency for massive datasets. Wang \& Ma (2021) investigate the optimal subsampling probability based on quantile regression, obtain the most subsampling probability for the two versions under quantile regression, and derive the asymptotic distribution based on the subsample estimates.  Yuan et al. (2022), Shao \&  Wang (2022) developed optimal subsampling for composite quantile regression with large datasets. Keret \& Gorfine (2023) developed the Cox regression subsampling-based estimators to enhance computational efficiency when dealing with right censored and possibly left-truncated data containing rare events.

The studies mentioned above mainly explore subsampling methods for statistical models with scalar variables, with relatively limited research conducted on subsampling for functional regression. Recently, Liu et al. (2023) proposed an optimal subsampling strategy for functional generalized linear models that follows the L-optimality criterion and establishes its asymptotic properties for the subsampling estimator. Yan et al. (2023) studied the optimal subsampling algorithm for quantile regression under functional data. So far as we know, there is no research on composite quantile regression for large-scale functional datasets, which motivates us to study the optimal subsampling algorithm for composite quantile regression on large-scale functional data.

The composite quantile regression estimator is robust to heavy-tailed errors. It is also robust to outliers in the responses. It offers improved estimation efficiency compared to single quantile regression. Previous research by Zou \& Yuan (2008) , Kai et al. (2010)  have shown that estimation methods based on composite quantile regression models have the potential to significantly outperform those based on ordinary least squares estimation. Jiang \& Sun (2022) investigated the application of composite quantile regression to single-index models in the context of ultra-high-dimensional data. Jiang \& Huang (2022) applied robust quantile regression to the Tecator dataset using a single-index partially functional linear model. Jiang et al. (2023) introduced a composite quantile regression estimator based on the B-spline basis function for functional single-indicator models under the influence of non-normal error.

For these reasons, we investigate optimal subsampling techniques based on composite quantile regression models in massive functional datasets. We provide an in-depth analysis of the asymptotic distribution of the subsampling estimator. We propose a new optimal subsampling algorithm based on the functional L-optimality criterion. The implementation of L-optimal subsampling probabilities is straightforward since they are independent of the response densities given covariates. Using these sampling probabilities, we introduce algorithms for computing and estimating subsamples and establish their asymptotic distributions and optimality. And based on numerical simulation studies and analysis of real data, we demonstrate that our proposed optimal subsampling method always outperforms the uniform subsampling method.

The rest of the paper is organized as follows: In Section 2, we give the composite quantile regression estimation of the slope function using the B-spline method and prove the asymptotic property of the composite quantile regression estimation and we give the optimal subsample probability under the L-optimality criterion by the asymptotic variance and give the two-step algorithm; In Section 3, we illustrate the method proposed in this paper by numerical simulation; In Section 4, we apply the method proposed in this paper on a real daily PM2.5 dataset to validate the method proposed in this paper.

\section{Methodology} \setcounter {equation}{0}
\def\theequation{\thesection.\arabic{equation}}

We consider a generalized linear model under data where the response variable is scalar and the covariates are of functional type
\begin{eqnarray}\label{model}
	y_i&=&\int_0^1x_i(t)\beta_0(t)dt +\varepsilon_i, i=1,\cdots,N,
\end{eqnarray}
 where $x_i(t)$ is an independent realization of the unknown process $x(t)$, $y_i$ is a continuous scaled response variable, and $\beta_0(t)$ is the slope function. Assume that the
random error $\varepsilon_i$ has cumulative distribution function $F(\cdot)$ and probability density function $f(\cdot)$.
Assume the cumulative distribution function and the probability density function of the random error $\varepsilon_i$ are $F(\cdot)$ and $f(\cdot)$, respectively.

\subsection{The Composite Quantile Regression Estimation of $\beta(t)$}

Let $K$ be a fixed composite level of CQR, independent of the sample size $N$. For this composite level, define the equally spaced quantile levels as $\tau_k=k/(K+1)$, $k=1,\cdots,K$. For  $\tau \in (0,1)$  and $\u\in\mathbb{R}$, let $\rho_\tau(u)=u\{\tau-I(u<0)\}$ represent the check loss function for the $\tau$-th quantile level. The coefficient function $\beta(t)$ in model (\ref{model}) is estimated by minimizing the following function:
\begin{eqnarray*}
	Q(\beta)=\sum_{i=1}^N \sum_{k=1}^K \rho_{\tau_k} \left(y_i-b_k- \int_0^1 x_i(t)\beta(t)dt  \right)
	+ \frac{\lambda}{2} \int_0^1 \left\{\beta^{(q)}(t)\right\}^2dt,
\end{eqnarray*}
where, $\beta^{(q)}(t)$  denotes the $q$th order derivative of $\beta(t)$ with $q\leq p$. We now briefly review the smoothing spline method for estimating $\beta(t)$. B-spline basis functions are defined by their order and a sequence of knots. Specifically, consider dividing the interval [0, 1] into $M + 1$ sub-intervals using $M$ equally spaced interior knots, denoted as $ [t_j, t_{j+1}]  for j=0,\cdots,M$. Within these sub-intervals, we can define $M + p + 1$ normalized B-spline basis functions$\{B_l(t), 1\leq l \leq M+p+1\}$, represented as

\begin{eqnarray*}
	\B(t) &=& (B_1(t),\cdots,B_{M+p+1}(t))^\textsf{T}.
\end{eqnarray*}
These basis functions are piecewise polynomials of degree $p$ on each subinterval $[t_j, t_{j+1}]$ and are $p-1$ times continuously differentiable on $[0, 1]$. Let $S_{Mp}$ denote the linear space spanned by the B-spline basis functions $\{B_l(t):l=1,\cdots,M+p+1\}$.

Since the minimization of $ Q(\beta)$ is over all $\beta \in S_{Mp} $, we represent $\beta(t)$ as a linear combination of B-spline basis functions:
\begin{eqnarray*}
	\beta(t) &=& \sum_{l=1}^{M+p+1} B_l(t) \theta_l(t)= \B^\textsf{T}(t) \vtheta.
\end{eqnarray*}

The estimator $\hat{\vtheta} $ of $\vtheta$ can be obtained by minimizing the following penalized function:
\begin{eqnarray} \label{CQRestFull}
	Q (\vtheta, \b, \lambda )&=& \sum_{i=1}^N \sum_{k=1}^K \rho_{\tau_k} \left(y_i-b_k- \left[\int_0^1 x_i(t)\B(t)dt \right] ^\textsf{T}\vtheta  \right)
	\nonumber\\
	&&+\frac{\lambda}{2} \int_0^1 \left\{\B^{(q)\textsf{T}}(t) \vtheta\right\}^2dt.
\end{eqnarray}

Let $\U_i = \int_0^1 x_i(t)\B(t)dt $ and $\D_q=\int_0^1\B^{(q)}(t)\B^{(q)\textsf{T}}(t)dt$. The function $Q (\vtheta, \b, \lambda )$ can then be rewritten as:

\begin{eqnarray} \label{CQRfunFull}
	Q (\vtheta, \b, \lambda )&=& \sum_{i=1}^N \sum_{k=1}^K \rho_{\tau_k} \left(y_i-b_k- \U_i  \vtheta  \right)+\frac{\lambda}{2} \vtheta^\textsf{T} \D_q \vtheta.
\end{eqnarray}

Let \ $(\hat{\vtheta}_{N},\hat{\b}_{N})=\arg\min_{\vtheta, \b} Q (\vtheta, \b, \lambda ).$ The estimator of $\beta(t)$ can be written as $\hat{\beta}(t)=\B^\textsf{T}(t)\hat{\vtheta}_{N}$.
For large  $N$, the asymptotic normality of the CQR estimator $\hat{\vtheta}_{N}$ forms the basis for statistical inference on $\vtheta_0$. However, it is computationally challenging to calculate $\hat{\vtheta}_{N}$ for massive datasets with very large $N$. In the following sections, we address these challenges by developing a subsampling approach to apply CQR to massive data.

\subsection{Subsampling based estimator and its asymptotic distribution}

In this subsection, we introduce a general subsampling algorithm and establish the asymptotic normality of the resulting estimator.
First, we take a random subsample with replacement from the full dataset according to probabilities  $\pi_i$, $i=1,\cdots, N$, where $\sum_{i=1}^N \pi_i=1$. These probabilities may depend on the observed data. Let  $R_i$ be the number of times the $i$-th data point is selected in the subsample, with $\sum_{i=1}^N R_i = n$. Since subsampling is with replacement, each $R_i$ follows a binomial distribution $Bin(n, \pi_i)$. Denote the full dataset by $\mathcal{D}_N=\{(x_i(t),y_i), i=1,\cdots,N, t \in [0,1]\}$ and the subsample by $\{(x_i^{*}(t),y_i^*,\pi_i^*), i=1,\cdots,n, t \in [0,1]\}$, where $x_i^*(t)$, $y_i^*$, and $\pi_i^*$ represent covariate functions, responses, and subsampling probabilities, respectively. The subsample-based CQR estimator $\hat{\vtheta}_S$ and $\hat{\b}_S$ are obtained by minimizing the following function:

\begin{eqnarray} \label{CQRestFull}
	Q^*(\vtheta, \b, \lambda )&=& \sum_{i=1}^n \sum_{k=1}^K \frac{1}{N\pi^*_i} \rho_{\tau_k} \left(y_i^*-b_k- \U_i^{*\textsf{T}}  \vtheta  \right)+\frac{\lambda}{2} \vtheta^\textsf{T} \D_q \vtheta
	\\
	&=&\sum_{i=1}^N \sum_{k=1}^K \frac{R_i }{N\pi_i} \rho_{\tau_k} \left(y_i-b_k- \U_i^{\textsf{T}}  \vtheta  \right)+\frac{\lambda}{2} \vtheta^\textsf{T} \D_q \vtheta,
	\nonumber
\end{eqnarray}
where  $\U_i = \int_0^1 x_i(t)\B(t)dt $. To describe the asymptotic form of $\hat{\vtheta}_S$, we need the following preparations. Let $b_{0k}=\inf\{u: F(u)\geq\tau_k\}$  for $k=1,\cdots,K$ and $\tilde{\U}_{ik}=(\U_i^\textsf{T}, \e_k^\textsf{T})^\textsf{T}$, where $\e_k$  is a $K \times 1$ vector with a one in its $k$-th position and zeros elsewhere. To establish the asymptotic normality of $\hat{\beta}_S(t)=\B^\textsf{T}(t)\hat{\vtheta}_S$, we assume the following conditions:

\begin{itemize}
\item[(C1)] For the functional covariate $x(t)$,  assume there exist a constant $c_1$ such that $\|x(t)\|_2 \leq c_1 < \infty$  a.s..
\item[(C2)] Assume the unknown functional coefficient $\beta(t)$ is sufficiently smooth. That is, $\beta(t)$ has a $d^\prime$-th derivative $\beta^{d^\prime}(t)(t)$ such that
	\begin{equation*}
		|\beta^{d^\prime}(t)(t)-\beta^{d^\prime}(t)(s) |\leq c_2 |t-s|^v, t,s \in [0,1],
	\end{equation*}
	where the constant $c_2>0$ and $v \in [0,1]$. In what follows, we set $d=d^\prime+v \geq p+1$.
\item[(C3)] Assume the density functions $f\left(\varepsilon_i\right), i=1, \cdots, N$, are continuous and uniformly bounded away from 0 and $\infty$ at $\varepsilon_i=0$. Furthermore, assume $\max _{i=1, \cdots, N} E\left(\varepsilon_i^4\right)<\infty$.
\item[(C4)] Assume the smoothing parameter $\lambda$ satisfies $\lambda=o(N^{1/2}M^{1/2-2q})$ with $q\leq p$.
\item [(C5)] Assume the number of knots $M=o(N^{1/2})$ and $M/N^{1/(2d+1)}\rightarrow \infty$ as $N\rightarrow \infty$.
\item [(C6)] Assume that $\max_{1\leq i \leq N} (N \pi_i)^{-1}=O(n^{-1}) $, $\max_{1\leq i \leq N} \{\|\U_i\|+1\}/\pi_i=o_p(\sqrt{n}N )$ and $n=o(M^2)$.
\end{itemize}
\begin{theorem}\label{subest}
	Under the Assumptions C1-C6,  for $t \in [0,1]$, as $n,N\rightarrow \infty$, we have
	\begin{equation*}
		\left\{\B^\textsf{T}(t)\H_N^{-1} \V_{\pi} \H_N^{-1} \B(t) \right\}^{-1/2} \sqrt{\frac{n}{M}} \left(\hat{\beta}_S(t)-\beta(t) \right)\drow N(0,1),
	\end{equation*}
	where
	\begin{eqnarray*}
		\V_{\pi}  &=&  \frac{1}{M} \sum_{i=1}^N \frac{1}{N^2 \pi_i} \biggr[\sum_{k=1}^K  \{I\left(\varepsilon_i<b_{0k}\right)-\tau_k\} \tilde{\U}_{ik}  \biggr]^{\otimes 2} \\
		\H_N&=&\G_N+\frac{\lambda}{N} \D_q,\\
		\G_N&=&\frac{1}{N}\sum_{i=1}^N \sum_{k=1}^K f\left(b_{0k}\right)  \tilde{\U}_{ik}^{\otimes 2},
	\end{eqnarray*}
	for a vector $\u$, $\u^{\otimes 2} =\u \u^ T$.
\end{theorem}
\subsection{Optimal subsampling probabilities}
In this subsection, we consider how to specify the subsampling distribution $\pi = \{\pi_i\}_{i=1}^n$ with theoretical backing. From Theorem \ref{subest}, it can be deduced that under certain conditions, $\hat{\beta}_S(t)$ is asymptotically unbiased. Our goal is to identify the optimal subsampling probabilities that minimize the asymptotic integrated mean squared error (IMSE) of $\hat{\beta}_S(t)$. The asymptotic IMSE is defined as follows:

\begin{eqnarray}\label{IMSE}
	\rm{IMSE}(\hat{\beta}_S) &=& \frac{M}{n} \int_0^1 \B^\textsf{T}(t)\H_N^{-1} \V_{\pi} \H_N^{-1} \B(t) dt.
\end{eqnarray}
In (\ref{IMSE}), the matrix $\H_N^{-1} \V_{\pi} \H_N^{-1}$ represents the asymptotic variance-covariance matrix of $\sqrt{n/M}(\hat{\vtheta})$. The inequality $\int_0^1 \B^\textsf{T}(t) \Delta \B(t) dt \leq \int_0^1 \B^\textsf{T}(t) \Delta^\prime \B(t) dt$ holds if and only if $\Delta \leq \Delta^\prime$ in the Lowner ordering sense. Therefore, our aim is to minimize the asymptotic variance-covariance matrix $\H_N^{-1} \V_{\pi} \H_N^{-1}$ by selecting the subsampling probabilities that minimize $tr(\H_N^{-1} \V_{\pi} \H_N^{-1})$. This objective aligns with the A-optimality criterion in optimal experimental design (refer to Atkinson et al., 2007). By following this criterion, we can derive an explicit formula for the optimal subsampling probabilities as presented in the following theorem.

\begin{theorem}\label{AoptTh}
	
	If the sampling probabilities $\pi_i$, $i=1,\cdots,N$, are chosen as
	\begin{eqnarray*}
		\pi_{i}^{Aopt} &=& \frac{\|\sum_{k=1}^K\{\tau_k-I(\varepsilon_i<b_{0k})\}\H_N^{-1} \tilde{\U} _{ik}\|}{\sum_{j=1}^N\|\sum_{k=1}^K\{\tau_k-I(\varepsilon_j<b_{0k})\}\H_N^{-1}\tilde{\U}_{jk}\| },\ \ i=1,\cdots,N,
	\end{eqnarray*}
	then $\tr(\H_N^{-1}\V_\pi \H_N^{-1}) $ attains its minimum.
	
\end{theorem}

The A-optimal subsampling probabilities $\pi_{i}^{Aopt}(i=1,\cdots,N)$ are based on the unknown density function $f(\cdot)$. By following Wang \& Ma (2020), we derive the optimal subsampling probabilities under the L-optimality criterion, which aims to minimize $tr(\V_\pi)$ with respect to $\pi$. It is evident that

\begin{theorem} \label{LoptTh} If the sampling probabilities $\pi_i$, $i=1,\cdots,N$, are chosen as
	\begin{eqnarray}\label{pilopt}
		\pi_{i}^{Lopt}= \frac{\|\sum_{k=1}^K\{\tau_k-I(\varepsilon_i<b_{0k})\}\tilde{\U}_{ik}\|}{\sum_{j=1}^N\|\sum_{k=1}^K\{\tau_k-I(\varepsilon_j<b_{0k})\}\tilde{\U}_{jk}\| },\ \ i=1,\cdots,N,
	\end{eqnarray}
	then $\tr(\V_\pi) $ attains its minimum.
\end{theorem}

Compared to the A-optimal subsampling probabilities $\pi_{i}^{Aopt}$, computing the L-optimal subsampling probabilities $\pi_{i}^{Lopt}$ is significantly simpler, as it doesn't require the computation of the unknown $\H_N$. Considering this advantage, our focus is solely on developing algorithms and statistical properties for the L-optimal subsampling probabilities.
In (\ref{pilopt}), notice that $\varepsilon_i=y_i-\U_i^\textsf{T}\vtheta_0$, where $i=1,\cdots,N$. Consequently, $\pi_{i}^{Lopt}$ is dependent on the unknown $\vtheta_0$ and $b_{0k}$. This dependency renders the direct implementation of the L-optimal weight result challenging. To address this issue, we propose a two-step algorithm as follows.


\paragraph{Algorithm 1} Two-step Algorithm in Implementing $\pi_{i}^{Lopt}$.
\begin{itemize}
	\item \textbf{Step 1:} Using the uniform sampling probabilities $\pi_i^U=1/N$, $i=1,\cdots,N$, draw a random subsample of size $n_0$, denoted by $S_0$, to obtain a preliminary estimate of $\vbeta_0 $, denoted by $\tilde{\vbeta}_U$. Replace $\vbeta_0$ with $\tilde{\vbeta}_U$ in (\ref{pilopt}) to obtain the approximate optimal subsampling probabilities
	$$\tilde{\pi}_{i}^{*Lopt}= \frac{\| \psi_{i}(\tilde{\vbeta}_U)\| }{\sum_{j=1}^N \| \psi_{j}(\tilde{\vbeta}_U)\|}.$$
	\item \textbf{Step 2:} Using $\{\tilde{\pi}_{i}^{Lopt}\}_{i=1}^N$, sample with replacement to obtain a subsample, \par $\{({\bf{x}}_i^{*\textsf{T}},y_i^*,\tilde{\pi}_{i}^{*Lopt}), i=1,\cdots,n\}$, and calculate the estimate of $\vbeta_0$ through
	\begin{eqnarray} \label{hatpivtheta}
		\hat{\vbeta}_{Lopt}=\arg\min_{\vbeta\in\Theta}n^{-2}\sum_{i=1}^n\sum_{j=1}^n \frac{1}{N^2\tilde{\pi}_{i}^{*Lopt}\tilde{\pi}_{j}^{*Lopt} } \left|e_i^*(\vbeta)-e_j^*(\vbeta) \right|,
	\end{eqnarray}
	where $e_i^*(\vbeta)=y^*- {\bf{x}}_i^{*\textsf{T}}\vbeta$. The minimizer $\tilde{\vbeta}_{Lopt}$ in (\ref{hatpivtheta}) is computed using the Barrodale and Roberts algorithm, implemented by the $rq$ function in the R package $quantreg$.
\end{itemize}

\section{Simulation studies}
\setcounter{equation}{0}\def\theequation{\thesection.\arabic{equation}}

In this section, we evaluate the performance of the subsample-based estimator using the two-step procedure outlined in Algorithm 1. We conduct 500 replications with datasets generated from the following functional linear model:
\begin{eqnarray}\label{generate Y}
	y_i &=& \int_0^1  x_i (t)\beta(t) dt+\sigma\varepsilon_i, i=1,\cdots,n,
\end{eqnarray}

where $\sigma=0.5$. The random errors $\varepsilon_i$'s are generated in two cases:
\par
\begin{itemize}
\item[(1)] $ N(0,1) $, the standard normal distribution.\par
\item[(2)] $ t_3 $, the $t$ distribution with 3 degree of freedom.\par
\end{itemize}

Similar to Liu et al. (2023), we generate the functional covariates as $$x_i(t) = \sum a_{ij}\B_j(t),$$ where $\B_j(t)$ represents the cubic B-spline basis functions that are sampled at 100 equally spaced points between 0 and 1. Let $\A=(a_{ij})$, we explore three distinct distributions for the basis coefficients $\a_{i}=(a_{i1},\cdots,a_{ij},\cdots)$:\par
\begin{itemize}
\item[(1)] $N(\textbf{0},\vSigma) $, the multivariate normal distribution with $\vSigma_{ij}=(0.5^{\mid i-j \mid})$.\par
\item[(2)] $t_2$, the multivariate $t$ distribution with 2 degree of freedom, $t_2(0,\vSigma)$.\par
\item[(3)] $t_3$, the multivariate $t$ distribution with 3 degree of freedom, $t_3(0,\vSigma)$.\\
\end{itemize}

Three forms of $\beta(t)$ are considered: \par
\begin{itemize}
\item[(1)] $\beta(t)=\sqrt{2} \{\cos (2\pi t)\}$, \par
\item[(2)] $\beta(t)=\exp\{-32 (t-0.5)^2\}+2t-\sqrt{2}\{\sin (2\pi t)+\cos (2\pi t)\}$,\par
\item[(3)] $\beta(t)=2t^2+0.25t+1$.\par
\end{itemize}

To compare the L-optimal subsampling (Lopt) method with the uniform subsampling (Unif) method, we compute the root integrated mean squared error (IMSE) from 1000 repetitions:
\begin{eqnarray*}
	\mbox{IMSE}=\frac{1}{1000} \sum_{k=1}^{1000} \sqrt{\int_0^1 \Big\{ \tilde{\beta}^{(k)} (t) - \beta(t) \Big\} ^2 dt},
\end{eqnarray*}
where $\tilde{\beta}^{(k)} (t)$ is the estimator from the $k$-th repetition.
To assess the performance of the proposed subsampling methods, we take $N=10^5$, $n \in \{ 600, 800, 1000, 1200, 1400, 1600 \}$.
The simulation results are displayed in Figures 1-3. It is clear to see that the Lopt subsampling method always have smaller IMSEs than the Unif subsampling method for all cases. This indicates that the method proposed in this article is more effective.

To further evaluate the computational efficiency of the subsampling methods, we record the computing time of the three subsampling methods. Since all the cases have similar performance, we only show the results when $\varepsilon_i\sim N(0,1)$, and $\a_{i} \sim N(\textbf{0},\vSigma) $. The results corresponding to the first two $\beta(t)$ on different $n$ for the Unif and Lopt subsampling methods with $K = 9$ and $N  = 10^5$ are given in Tables 1-3. It is not surprising that the Unif method takes the least amount of time since it does not require calculating additional optimal subsampling probabilities. The computation time of the full data, as shown in the last row of Tables 1-3, further confirms that our proposed methods are effective in reducing the computational burden.

\begin{figure}
	\centering
	\includegraphics[width=1\linewidth]{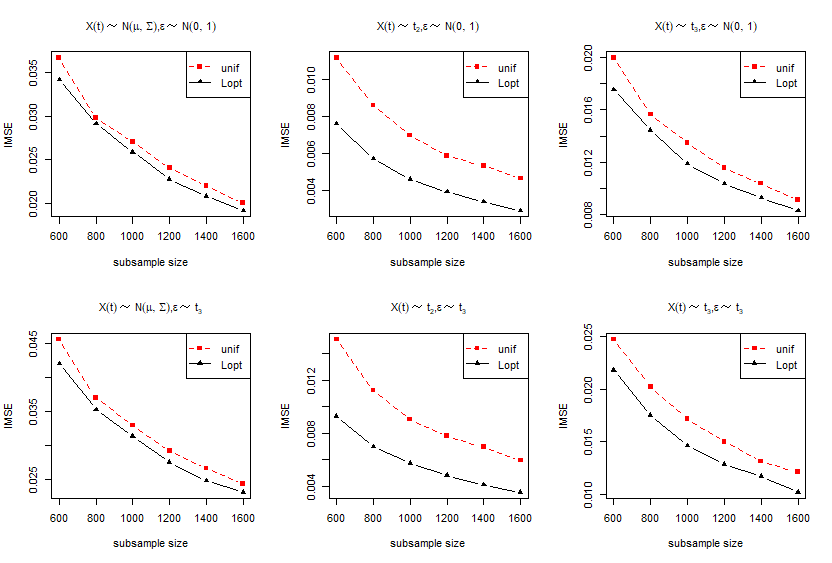}
	\caption{IMSE for different subsampling size $n$ when three distributions when  and $N=10^5$.}\label{mnf1}
\end{figure}

\begin{figure}
	\centering
	\includegraphics[width=1\linewidth]{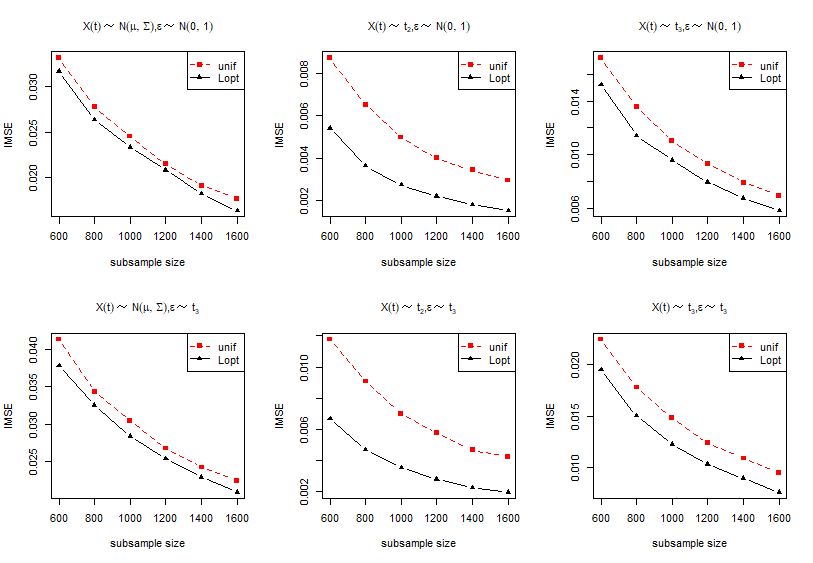}
	\caption{IMSE for different subsampling size $r$ when three distributions when $\beta(t)=\exp\{-32(t-0.5)^2\}+2t-\sqrt{2}\{\sin(2\pi t)+\cos(2\pi t)\}$ and $N=10^5$.}\label{mnf2}
\end{figure}

\begin{figure}
	\centering
	\includegraphics[width=1\linewidth]{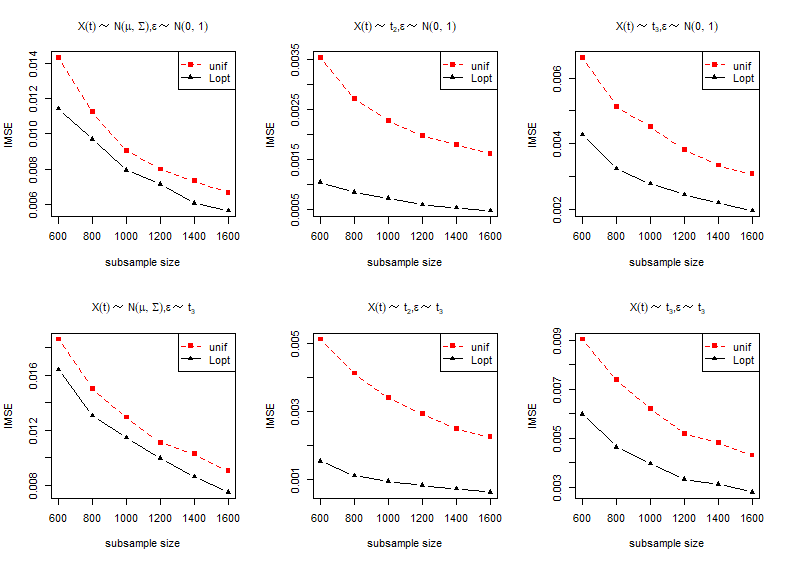}
	\caption{IMSE for different subsampling size $r$ when three distributions when $\beta(t)=2t^2+0.25t+1$ and $N=10^5$.}\label{mnf3}
\end{figure}

\begin{table*}[!t]
	\centering
	\caption{CPU seconds for different subsampling size $r$ with $K=9$ and $N=10^5$ for 1000 repetitions when $\beta(t)=\sqrt{2} \{\cos(2\pi t)\}$.\label{tab:1}}
	\begin{tabular*}{\textwidth}{@{\extracolsep\fill}lccccccl@{\extracolsep\fill}}
		\toprule
		& \multicolumn{6}{c}{$r$} & \\
		\cmidrule{2-7}
		method & 600 & 800 & 1000 & 1200 & 1400 & 1600 & \\
		\midrule
		Unif   & 0.051 & 0.054 & 0.055 & 0.061 & 0.076 & 0.093 & \\
		Lopt   & 0.064 & 0.064 & 0.067 & 0.073 & 0.087 & 0.102 & \\
		\midrule
		Full data CPU seconds & 14.307 & & & & & & \\
		\bottomrule
	\end{tabular*}
\end{table*}

\begin{table*}[!t]
	\centering
	\caption{CPU seconds for different subsampling size $r$ with $K=9$ and $N=10^5$ for 1000 repetitions when $\beta(t)=\exp\{-32(t-0.5)^2\}+2t-\sqrt{2}\{\sin(2\pi t)+\cos(2\pi t)\}$.\label{tab:2}}
	\begin{tabular*}{\textwidth}{@{\extracolsep\fill}lccccccl@{\extracolsep\fill}}
		\toprule
		& \multicolumn{6}{c}{$r$} & \\
		\cmidrule(lr){2-7}
		method & 600 & 800 & 1000 & 1200 & 1400 & 1600 & \\
		\midrule
		Unif   & 0.066 & 0.084 & 0.101 & 0.123 & 0.141 & 0.142 & \\
		Lopt   & 0.079 & 0.100 & 0.115 & 0.137 & 0.159 & 0.158 & \\
		\midrule
		Full data CPU seconds
		& 14.134 & & & & & & \\
		\bottomrule
	\end{tabular*}
\end{table*}

\begin{table*}[!t]
	\centering
	\caption{CPU seconds for different subsampling size $r$ with $K=9$ and $N=10^5$ for 1000 repetitions when $\beta(t)=2t^2+0.25t+1$.\label{tab:3}}
	\begin{tabular*}{\textwidth}{@{\extracolsep\fill}lccccccl@{\extracolsep\fill}}
		\toprule
		& \multicolumn{6}{c}{$r$} & \\
		\cmidrule(lr){2-7}
		method & 600 & 800 & 1000 & 1200 & 1400 & 1600 & \\
		\midrule
		Unif   & 0.052 & 0.054 & 0.059 & 0.061 & 0.069 & 0.078 & \\
		Lopt   & 0.072 & 0.076 & 0.084 & 0.083 & 0.091 & 0.097 & \\
		\midrule
		Full data CPU seconds & 14.396 & & & & & & \\
		\bottomrule
	\end{tabular*}
\end{table*}

\section{An application}
\setcounter {equation}{0}
\def\theequation{\thesection.\arabic{equation}}
The concentration of PM2.5 is an important indicator of air quality, which refers to particulate matter with a diameter equal to or smaller than 2.5 $\mu m$  in the air. Its association with respiratory inflammation and other breathing complications is extensively documented in studies such as those by Wang et al.(2016) and Zhang et al.(2015).
In this section, we illustrate the proposed method on a PM2.5 dataset (https://archive-beta.ics.uci.edu/ml/
datasets/beijing+multi+site+air+quality+data), which encompasses hourly air pollution data from 12 national control air quality monitoring sites in Beijing. The data spans from March 1, 2013, to February 28, 2017.
Our primary objective is to use a trajectory of CO ($mg/m^{3}$) over a span of 23 hours to predict the maximum PM2.5 concentration ($mg/m^{3}$) for the same day.
After removing missing values, we have a dataset of 13,530 days with complete records. Since the true value of $\beta(t)$ is unknown, we take the estimator from the full data instead. We calculate the empirical IMSE (eIMSE) using
$$\rm{eIMSE}=\frac{1}{1000} \sum_{k=1}^{1000} \sqrt{\int_0^1 \Big\{ \tilde{\beta}^{(k)} (t) - \hat{\beta}(t) \Big\} ^2 dt}$$,
and compare the Lopt subsampling method with uniform subsampling method.

The results in Figure 4 demonstrate that under composite level K=15 and K=9, the L-optimal subsampling method consistently outperforms the uniform subsampling method by having lower eIMSE values. As $\textit{n}$ increases, all eIMSE values decrease, demostrate the consistent estimation properties of the sub-sampling method.

\begin{figure}
	\centering
	\includegraphics[width=1\linewidth]{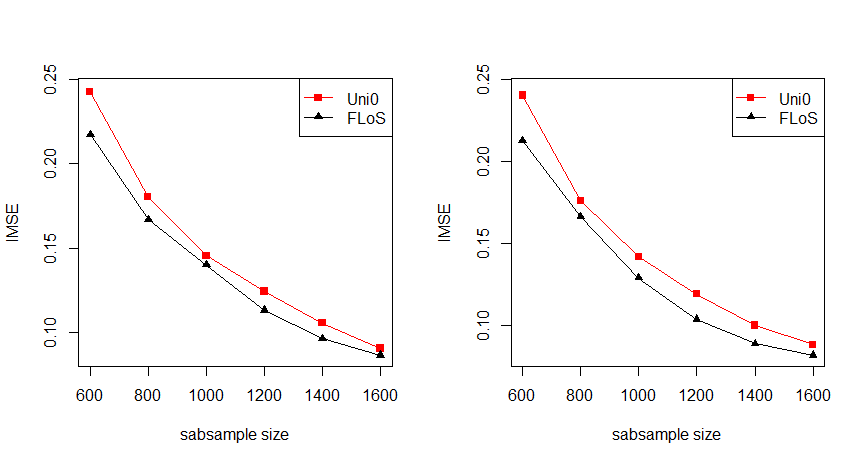}
	\caption{IMSE for different subsampling size $r$ when K=9(left) and K=15(right)}\label{SLQ=915}
\end{figure}

\section*{Acknowledgements}

Xiaohui Yuan was partly supported by the National Social Science Fund of China (22BTJ019) and  Scientific Research Project of Jilin Provincial Department of Education (JJKH20230749KJ).


\makeatletter
\renewenvironment{thebibliography}[1]
{\section*{\refname}%
\@mkboth{\MakeUppercase\refname}{\MakeUppercase\refname}%
\list{\@biblabel{\@arabic\c@enumiv}}%
{\settowidth\labelwidth{\@biblabel{#1}}%
\leftmargin\labelwidth \advance\leftmargin\labelsep
\advance\leftmargin by 2em%
\itemindent -2em%
\@openbib@code
\usecounter{enumiv}%
\let\p@enumiv\@empty
\renewcommand\theenumiv{\@arabic\c@enumiv}}%
\sloppy \clubpenalty4000 \@clubpenalty \clubpenalty
\widowpenalty4000%
\sfcode`\.\@m} {\def\@noitemerr
{\@latex@warning{Empty `thebibliography' environment}}%
\endlist}
\renewcommand\@biblabel[1]{}
\makeatother
\appendix
\section*{Appendix}
\newcounter{lemm}
\newtheorem{lemm}{Lemma A.}
\setcounter {equation}{0}
\def\theequation{A.\arabic{equation}}

To prove our theorems, we begin with the following several lemmas.


\begin{lemm}\label{RateG}
	Under Assumption C1 and C5, for any vector $\vmu \in \mathbb{R}^{M+p+1}$, there are some positive constants $c_3$, $c_4$, $c_5$ and $c_6$ such that
	\begin{eqnarray*}
		&&c_3M^{-1}\leq \sigma_{min}(\G_N) \leq \sigma_{max}(\G_N) \leq c_4 M^{-1},  \\
		&&c_5M^{2p-1}\|\vmu\|_2^2 \leq \vmu^T \D_q \vmu \leq c_6 M^{2q-1}  \|\vmu\|_2^2,
	\end{eqnarray*}
	where $\sigma_{min}(\cdot)$, $\sigma_{max}(\cdot)$ denote the smallest and largest eigenvalues of a matrix, respectively. In addition, we have $\|\G_N\|_{\infty}=O(M^{-1})$ and $\|\D_q\|_{\infty}=O(M^{2q-1})$.
\end{lemm}
\textbf{\emph{The proof of Lemma \ref{RateG}.}} These results can be derived directly from Lemma 7 and 8 in Liu et al.(2023).


\begin{lemm}\label{RateH}
	Under Assumption C1, and C3-C5, there are two positive constants $c_7$ and $c_8$ such that
	\begin{eqnarray*}
		&&c_7M^{-1}\leq \sigma_{min}(\H_N) \leq \sigma_{max}(\H_N) \leq c_8 M^{-1},
	\end{eqnarray*}
	and $\|\H_N\|_{\infty}=O(M^{-1})$.
\end{lemm}
\textbf{\emph{The proof of Lemma \ref{RateH}.}} These results can be derived directly from Lemma 9 in Liu et al.(2023).

\begin{lemm}\label{RatesplineBeta} Let
	\begin{equation*}
		b_a(t)=-\frac{\beta^d(t)}{M^d d!}\sum_{j=0}^M I(t_j\leq t<t_{j+1})Br_d\left( \frac{t-t_j}{M^{-1}}\right)  =O(M^{-d})
	\end{equation*}
	be the spline approximation bias and $Br_d(t)$ be the $d$-th Bernoulli polynomial.
	Under Assumption C2,  $$\B^\textsf{T}(t)\vtheta-\beta(t)=b_a(t)+o(M^{-d}).$$
\end{lemm}
\textbf{\emph{The proof of Lemma \ref{RatesplineBeta}.}} The proof of this lemma can be found in Barrow and Smith (1978), and Zhou et al. (1998) .


\begin{lemm}\label{RateI}
	Let $\vdelta=(\vtheta^\textsf{T},\b^\textsf{T})^\textsf{T}$, $\tilde{\U}_{ik}=(\U_i^{\textsf{T}}, \e_k^\textsf{T})^\textsf{T}$,$\psi_\tau(u)=\tau-I(u<0)$ and $u_{ik}=\varepsilon_i-b_{0k}$. Under assumptions C1-C6, for any $\vdelta \in \mathbb{R}^{M+p+1}$, we have
	\begin{equation}
		-\sqrt{\frac{M}{n}} \sum_{i=1}^N \sum_{k=1}^K \frac{R_i}{N \pi_i} \vdelta^T \tilde{\U}_{i k} \psi_{\tau_k}\left(u_{ik}\right)=-\sqrt{M} {\mathbf{W}}^T \vdelta+o_p(1),\label{normalZ}
	\end{equation}
	where $  \bf{V}_\pi  ^{-\frac{1}{2}} {\bf{W}} \rightarrow N(\0, \I)$ in distribution.
\end{lemm}

\noindent\textbf{\emph{The proof of Lemma \ref{RateI}.}} Set
\begin{equation}
	Z_n=-\sqrt{\frac{M}{n}} \sum_{i=1}^N \sum_{k=1}^K \frac{R_i}{N \pi_i} \vdelta^T \tilde{\U}_{i k} \psi_{\tau_k}\left(u_{ik}\right)\nonumber
\end{equation}

Direct calculation shows that conditionally on $\mathcal{D}_N$,
\begin{eqnarray*}
	E\left(Z_n \mid \mathcal{D}_N\right) &=&-\sqrt{\frac{M}{n}} \sum_{i=1}^N \sum_{k=1}^K E\left[\left.\frac{R_i}{N \pi_i} \vdelta^T \tilde{\U}_{i k} \psi_{\tau_k}\left(u_{ik}\right) \right\rvert\, \mathcal{D}_N\right] \\
	&=&-\frac{\sqrt{M n}}{N} \sum_{i=1}^N \sum_{k=1}^K \vdelta^T \tilde{\U}_{ik} \psi_{\tau_k}\left(u_{ik}\right),
\end{eqnarray*}
\begin{eqnarray*}
	\mbox{Var}\left(Z_n \mid \mathcal{D}_N\right)&= & \sum_{i=1}^N \sum_{k=1}^K \frac{M}{N^2} \frac{1}{\pi_i}\left(\vdelta^T \tilde{\U}_{i k}\right)^2 \psi_{\tau_k}^2\left(u_{ik}\right)- \\
	&& \sum_{i=1}^N \sum_{j=1}^N \sum_{k=1}^K \frac{M}{N^2}\left[\vdelta^T \tilde{\U}_{i k} \psi_{\tau_k}\left(u_{ik}\right)\right]^T\left[\vdelta^T \tilde{\U}_{j k} \psi_{\tau_k}\left(u_{jk}\right)\right] \\
	&= & \frac{M}{N^2}\sum_{i=1}^N \sum_{k=1}^K \frac{1-\pi_i}{\pi_i}\left(\vdelta^T \tilde{\U}_{i k}\right)^2 \psi_{\tau_k}^2\left(u_{ik}\right)-o_p(1).
\end{eqnarray*}
By the fact that $P\left(y_i<b_{0k}+\int_0^1 x_i(t) \beta(t) d t \mid x_i(t)\right)=\tau_k$, we have
\begin{eqnarray*}
	&& E\left(\psi_{\tau_k}\left(u_{ik}\right) \mid x_i(t)\right) \\
	&=& \tau_k-E\left[I\left(u_{ik}<0\right) \mid x_i(t)\right] \\
	&=& P\left(y_i<b_{0k}+\int_0^1 x_i(t) \beta(t) d t \mid x_i(t)\right)-P\left(\varepsilon_i<b_{0 k} \mid x_i(t)\right) \\
	&=& P\left(y_i<b_{0k}+\int_0^1 x_i(t)\left(\beta_0(t)+b_a(t)(1+o p(1))\right) d t \mid x_i(t)\right) \\
	&& -P\left(\varepsilon_i<b_{0k} \mid x_i(t)\right) \\
	&=& b_i f\left(b_{0k}\right)(1+op(1))\\
	&=& o_p(1),
\end{eqnarray*}
where $b_i=\int_0^1 x_i(t) b_a(t) d t$,  and the fourth equality is obtained by the Taylor expansion of the cumulative distribution function of the error $\varepsilon_i $ at point $b_{0k}$. As a result, the conditional expectation of
$Z_n$ can be calculated as
\begin{eqnarray*}
	E\left(Z_n \right)
	&=&E\left[E\left(Z_n \mid \mathcal{D}_N\right)  \right] \\
	&=&\frac{\sqrt{M n}}{N} \sum_{i=1}^N \sum_{k=1}^K \vdelta^T \tilde{\U}_{i k} b_i f\left(b_{0k}\right)(1+o_p(1)).
\end{eqnarray*}
Since $x_i(t)$ are square integrable functions,   for $1\leq
l \leq M+p+1$, by the Cauchy-Schwarz inequality in integral form, there exist constant $c$ such that
\begin{eqnarray*}
	U_{il}^2   \leq   \int_0^1 x_i^2(t) d t \cdot \int_0^1 B_l^2(t) d t
	\leq c \int_0^1 B_l^2(t) d t,
\end{eqnarray*}
and
\begin{equation*}
	b_i^2 \leq c \int_0^1 b_a^2(t) d t.
\end{equation*}
By the property of B-spline function, $\int_0^1 \B(t) d t  =O\left(M^{-1}\right)$, and $b_a(t) =O\left(M^{-d}\right)$, we can find that $\left\|\U_{i}\right\|_{\infty}  =O\left(M^{-1}\right)$ and $b_i =O\left(M^{-d}\right)$. Putting them together, we have
$E\left(Z_n  \right)=O_p\left(\sqrt{M n} M^{-1-d}\right).$
On the other hand, according to law of total variance, the unconditional variance is given by
\begin{equation} \label{TotalVar}
	\operatorname{Var}\left(Z_n\right)=E\left[\operatorname{Var}\left(Z_n \mid \mathcal{D}_N\right)\right]+\operatorname{Var}\left[E\left(Z_n \mid \mathcal{D}_N\right)\right].
\end{equation}
For the first term in (\ref{TotalVar}), we get
\begin{eqnarray*}
	E\left[\operatorname{Var}\left(Z_n \mid \mathcal{D}_N\right)\right]
	& = & \frac{M}{N^2} E\left[\sum_{i=1}^N \sum_{k=1}^K \frac{1-\pi_i}{\pi_i}\left(\vdelta^T \tilde{\U}_{i k}\right)^2 \psi_{\tau_k}^2 (u_{ik}) \right]-o_p(1) \\
	&= & M  \vdelta^T\left[ \frac{1}{N^2} \sum_{i=1}^N \sum_{k=1}^K \left(\frac{1}{  \pi_i}- 1\right)\tilde{\U}_{i k} ^{\otimes 2} E\psi_{\tau_k}^2(u_{ik}) \right] \vdelta-o_p(1).
\end{eqnarray*}
The second term in (\ref{TotalVar}) equals
\begin{eqnarray*}
	\mbox{Var}\left[E\left(Z_n \mid \mathcal{D}_N\right)\right]
	&= & \frac{Mn}{N^2} \operatorname{Var}\left[\sum_{i=1}^N \sum_{k=1}^K \vdelta^T \tilde{\U}_{i k} \psi_{\tau_k}(u_{ik})\right] \\
	&=& \frac{M n}{N^2} \vdelta^T \sum_{i=1}^N \sum_{k=1}^K \tilde{\U}_{i k}^{\otimes 2}  E\psi_{\tau_k}^2(u_{ik})\vdelta-o_p(1).
\end{eqnarray*}
Thus, we have
\begin{eqnarray*}
	\operatorname{Var}\left(Z_n\right) & =&M \vdelta^T\left[\sum_{k=1}^K\sum_{i=1}^N  E\psi_{\tau_k}^2(u_{ik})\left(\frac{1}{N^2 \pi_i}-\frac{n-1}{N^2}  \right)\tilde{\U}_{i k} ^{\otimes 2}\right] \vdelta+o_p(1) \\
	&=&M \vdelta^T \V_\pi \vdelta+o_p(1).
\end{eqnarray*}
Denote $\varsigma_i=-\sqrt{\frac{M}{n}} \frac{R_i}{N \pi_i}\sum_{k=1}^K \vdelta^T \tilde{\U}_{i k} \psi_{\tau_k}(u_{ik})$.
We now check the Lindeberg-Feller conditions (Theorem 2.27 of van der Vaart 1998) under
the conditional distribution given $\mathcal{D}_N$.
For every $\varepsilon>0$,
\begin{eqnarray*}
	&&\sum_{i=1}^N E\left\{\left\|\varsigma_i\right\|^2 I\left(\left\|\varsigma_i\right\|>\varepsilon\right)\right\}
	\\
	&\leq & \frac{1}{\varepsilon} \sum_{i=1}^N E\left\{\left\|\varsigma_i\right\|^3\right\} \\
	&\leq & \left(\frac{M}{N}\right)^{\frac{3}{2}} \frac{1}{\varepsilon} \sum_{i=1}^N \sum_{k=1}^K \frac{E\left(R_i^3\right)}{N^3 \pi_i^3}|\vdelta^T \tilde{\U}_{i k} |^3 E\left[ |\psi_{\tau_k}(u_{ik})|^3 \mid x_i(t)\right] \\
	&= & o_p(1),
\end{eqnarray*}
where
\begin{equation*}
	E\left(R_i^3\right)=n(n-1)(n-2) \pi_i^3+3 n(n-1) \pi_i^2+n \pi_i,
\end{equation*}
and the last equality holds by combining Assumption C6 and the fact that $\left|\psi_{\tau_k}\left(u\right)\right| \leq 1$. Thus, by Lindeberg-Feller central limit theorem, it can be concluded that as $n \rightarrow \infty$, $N \rightarrow \infty$,
\begin{equation}
	\frac{Z_n-E\left(Z_n\right)}{\sqrt{\operatorname{Var}\left(Z_n\right)}} \rightarrow N(0,1)\nonumber,
\end{equation}
in distribution, which implies that the equation (\ref{normalZ}) holds because $E\left(Z_n\right)=O\left(\sqrt{M n} M^{-1-d}\right)=o_p(1)$. This finishes the proof of Lemma \ref{RateI}.\\


\noindent\begin{lemm}\label{RateJ}
	Let $v_{ik}=\sqrt{M / n} \vdelta^T \tilde{\U}_{i k}$.  Under the same assumptions in  Theorem \ref{subest}, we have
	\begin{equation}
		\sum_{i=1}^N \sum_{k=1}^K \frac{R_i}{N \pi_i} \int_0^{v_{ik}}\left[I\left(u_{ik} \leq s\right)-I\left(u_{ik} \leq 0\right)\right] d s=\frac{M}{2} \vdelta^T \G_N \vdelta+o_p(1).
	\end{equation}
\end{lemm}

\noindent\textbf{\emph{The proof of Lemma \ref{RateJ}.}} Let
\begin{equation}
	A_{n 2}(\vdelta)=\sum_{i=1}^N \sum_{k=1}^K \frac{R_i}{ \pi_i} \int_0^{v_{ik}}\left[I\left(u_{ik} \leq s\right)-I\left(u_{ik} \leq 0\right)\right] ds
\end{equation}
Observe that
\begin{eqnarray*}
	&& E\left\{ \frac{R_i }{N \pi_i} \int_0^{v_{ik}}\left[I\left(u_{ik} \leq s\right)-I\left(u_{ik} \leq 0\right)\right] d s  \right\} \\
	&=& E\left\{ \frac{E (R_i)}{N \pi_i}  \int_0^{v_{ik}}\left[I\left(u_{ik} \leq s\right)-I\left(u_{ik} \leq 0\right)\right] d s   \right\}  \\
	&= & \frac{n}{N} E\left\{\int_0^{v_{ik}}\left[I\left(u_{ik} \leq s\right)-I\left(u_{ik} \leq 0\right)\right] d s  \right\} \\
	&= & \frac{n}{N} \int_0^{v_{ik}}\left[P\left(\varepsilon_i \leq s+b_{0k} \mid x_i(t)\right)-P\left(\varepsilon_i \leq b_{0k}  \mid x_i(t)\right)\right] d s \\
	&= & \frac{\sqrt{M n}}{N} \int_0^{\vdelta^T \tilde{\U}_{i k}}\left[P\left(\varepsilon_i \leq l \sqrt{M / n}+b_{0k} \mid x_i(t)\right)-P\left(\varepsilon_i \leq b_{0k} \mid x_i(t)\right)\right] d l \\
	&= & \frac{M}{N} \int_0^{\vdelta^T \tilde{\U}_{i k}} f\left(b_{0k}\right) l d l \times (1+o_p(1)) \\
	&= & \frac{M}{2 N} f\left(b_{0k}\right)\left(\vdelta^T \tilde{\U}_{i k}\right)^2(1+o_p(1)).
\end{eqnarray*}
The total expectation of $E_n$ is given by

\begin{eqnarray*}
	E\left(A_{n 2}(\vdelta)/N\right) &=& \frac{M}{2 N} \sum_{i=1}^N \sum_{k=1}^K f\left(b_{0k}\right)\left(\vdelta^T \tilde{\U}_{i k}\right)^2(1+o_p(1)) \\
	&=& \frac{M}{2} \vdelta^T\left[\frac{1}{N} \sum_{i=1}^N \sum_{k=1}^K f\left(b_{0k}\right) \tilde{\U}_{i k} ^{\otimes 2}\right] \vdelta(1+o_p(1)) \\
	&=& \frac{M}{2} \vdelta^T \G_N \vdelta(1+o_p(1)).	
\end{eqnarray*}
Now, we show the total variance of $E_n$ satisfying $\operatorname{Var}\left(E_n\right)=o_p(1)$.
\begin{eqnarray*}
	\mbox{Var}\left(A_{n 2}(\vdelta)/N\right) &\leq&  E\left\{\sum_{i=1}^N \sum_{k=1}^K\frac{R_i}{N \pi_i} \int_0^{v_{ik}}\left[I\left(u_{ik} \leq s\right)-I\left(u_{ik} \leq 0\right)\right] ds\right\}^2 \\
	&\leq&  \sqrt{\frac{M}{n}}\left\{\max _{1\leq i \leq N} \frac{1}{N \pi_i}\left|\vdelta^T \tilde{\U}_{i k}\right|\right\} E\left(A_{n 2}(\vdelta)\right) \\
	&\leq&  \sqrt{\frac{M}{n}}\left\{\max _{1\leq i \leq N} \frac{1}{N \pi_i}\right\}\left\{\max _{1\leq i \leq N}\left|\vdelta^T \tilde{\U}_{i k}\right|\right\} E\left(A_{n 2}(\vdelta)/N\right),
\end{eqnarray*}
where the second inequality is from the fact that
\begin{eqnarray*}
	\int_0^{v_{ik}}\left[I\left(u_{ik} \leq s\right)-I\left(u_{ik} \leq 0\right)\right] ds
	&\leq& \int_0^{v_{ik}} \left|I\left(u_{ik} \leq s\right)-I\left(u_{ik} \leq 0\right)\right|ds \\
	&\leq&  \sqrt{\frac{M}{n}}\left|\vdelta^T \tilde{\U}_{i k}\right|, i=1,2, \cdots, N\nonumber.
\end{eqnarray*}
Thus, we have $E\left(A_{n 2}(\vdelta)/N\right)=O\left(M^{-1}\right)$ and $\operatorname{Var}\left(A_{n 2}(\vdelta)/N\right)=o_p(1)$. As a result, Lemma \ref{RateJ} holds by Chebyshev's inequality.\\


\bmsection{The proof of Theorem\label{app2}}
\noindent\textbf{\emph{The proof of Theorem \ref{subest} } }Let
\begin{eqnarray*}
	A_n(\vdelta) &=&  \sum_{i=1}^N \sum_{k=1}^K \frac{R_i}{\pi_i}\left[\rho_{\tau_k}\left(u_{ik}-v_{ik}\right)-\rho\left(u_{ik}\right)\right] \\
	&& +\frac{n \lambda}{2}\left(\vtheta_0+\sqrt{\frac{M}{n}} \vdelta\right)^T \D_q\left(\vtheta_0+\sqrt{\frac{M}{n}} \vdelta\right)-\frac{n\lambda}{2} \vtheta_0^T \D_q \vtheta_0\nonumber
\end{eqnarray*}
where $u_{ik}=\varepsilon_i-b_{0k}$ and $v_{ik}=\sqrt{M / n} \vdelta^T \tilde{\U}_{ik}$. It is easy to see that this function is convex and minimized at $\sqrt{n / M}\left(\hat{\vtheta}_S-\vtheta_0\right)$. Using Knight's identity(Knight 1998),
\begin{equation*}
	\rho_\tau(u-v)-\rho_\tau(u)=-v \psi_\tau(u)+\int_0^v[I(u \leq s)-I(u \leq 0)] ds,
\end{equation*}
we have
\begin{equation} \label{A_n}
	A_n(\vdelta)=A_{n 1}(\vdelta)+A_{n 2}(\vdelta)+A_{n 3}(\vdelta)+A_{n 4}(\vdelta),
\end{equation}
where
\begin{eqnarray*}
	&& A_{n 1}(\vdelta) = -\sqrt{\frac{M}{n}} \sum_{i=1}^N \sum_{k=1}^K \frac{R_i}{\pi_i} \vdelta^T \tilde{\U}_{i k} \psi_{\tau_k}\left(u_{ik}\right), \\
	&& A_{n 2}(\vdelta)=\sum_{i=1}^N \sum_{k=1}^K \frac{R_i }{\pi_i}\int_0^{v_{ik}}\left[I\left(u_{ik} \leq s\right)-I\left(u_{ik} \leq 0\right)\right] d s ,\\
	&& A_{n 3}(\vdelta)=\frac{M \lambda}{2} \vdelta^T \D_q \vdelta ,\\
	&& A_{n 4}(\vdelta)=\sqrt{M n} \lambda \vtheta_0^T \D_q \vdelta\nonumber.
\end{eqnarray*}
By Lemma \ref{RateI}, $A_{n 1}(\vdelta)$ satisfies
\begin{equation} \label{A_n1}
	\frac{A_{n 1}(\vdelta)}{N}=-\sqrt{M} {\bf{W}}^T \vdelta+o_p(1),
\end{equation}
where $  \bf{V}_\pi  ^{-\frac{1}{2}} {\bf{W}} \rightarrow N(\0, \I)$ in distribution. By Lemma \ref{RateJ}, we have
\begin{eqnarray} \label{A_n2}
	\frac{A_{n 2}(\vdelta)}{N}+\frac{A_{n 3}(\vdelta)}{N}  &=& \frac{M}{2} \vdelta^T\left(\G_N+\frac{\lambda}{N} \D_q\right) \vdelta+o_p(1) \notag\\
	&=& \frac{M}{2} \vdelta^T \H_N \vdelta+o_p(1).
\end{eqnarray}
From (\ref{A_n}), (\ref{A_n1}) and (\ref{A_n2}), we can obtain
\begin{equation}
	\frac{A_n(\vdelta)}{N}=-\sqrt{M} {\bf{W}}^T \vdelta+\frac{M}{2} \vdelta^T \H_N \vdelta+\frac{\sqrt{M n}}{N} \lambda \vtheta_0^T \D_q \vdelta+o_p(1)\nonumber.
\end{equation}
Since $A_n(\vdelta) / N$ is convex with respect to $\vdelta$ and has unique minimizer, from the corollary  of Hjort and Pollard (2011), its minimizer,   $\sqrt{n / M}\left(\hat{\vtheta}_S-\vtheta_0\right)$ satisfies
\begin{equation}
	\sqrt{\frac{n}{M}}\left(\hat{\vtheta}_S-\vtheta_0\right)=\H_N^{-1}\left(\frac{1}{\sqrt{M}} {\bf{W}}-\sqrt{\frac{n}{M}} \frac{\lambda}{N} \D_q \vtheta_0\right)+o_p(1)\nonumber.
\end{equation}
Because the random vector is only ${\bf{W}}$ in asymptotic form of $\hat{\vtheta}_S$ and $\hat{\beta}_S(t)-\beta_0(t)=\B^T(t)\left(\hat{\vtheta}_S-\vtheta_0\right)$, the expectation of $\hat{\beta}_S(t)-\beta_0(t)$ can be written as
\begin{equation}
	E\left[\hat{\beta}_S(t)-\beta_0(t)\right]=b_\lambda(t)(1+o_ p(1))\nonumber,
\end{equation}
where $b_\lambda(t)=-\frac{\lambda}{N} \B^T(t) \H_N^{-1} \D_q \vtheta_0$.

Let $\beta_0(t)=\B^T(t) \vtheta_0$. By Lemma \ref{RatesplineBeta}, the penalized spline estimator can be decomposed as
\begin{equation*}
	\hat{\beta}_S(t)-\beta(t)=\hat{\beta}_S(t)-\beta_0(t)+\beta_0(t)-\beta(t)= \hat{\beta}_S(t)-\beta_0(t)+b_a(t) +o(M^{-d}).
\end{equation*}
Then we have the asymptotic bias of $\hat{\beta}_S(t)$ as
\begin{equation}
	E\left[\hat{\beta}_S(t)-\beta(t)\right]=b_a(t)(1+o_ p(1))+b_\lambda(t)(1+o_p(1))\nonumber.
\end{equation}
Thus,
\begin{eqnarray*}
	&&\left\{\B^T(t) \H_N^{-1} \V_\pi \H_N^{-1} \B(t)\right\}^{-\frac{1}{2}} \sqrt{\frac{n}{M}}\left[\hat{\beta}_S(t)-\beta(t)-b_a(t)-b_\lambda(t)\right] \\
	&=&\left\{\B^T(t) \H_N^{-1} \V_\pi \H_N^{-1} \B(t)\right\}^{-\frac{1}{2}} \B^T(t) \H_N^{-1} \frac{1}{\sqrt{M}} {\bf{W}}+o_p(1)\nonumber.
\end{eqnarray*}
By the definition of ${\bf{W}}$ and Slutsky's Theorem, we can obtain for $t \in [0,1]$, as $N\rightarrow \infty$,
\begin{equation}
	\left\{\B^\textsf{T}(t)\H_N^{-1} \V_{\pi} \H_N^{-1} \B(t) \right\}^{-1/2}\sqrt{\frac{n}{M}} \left(\hat{\beta}_S(t)-\beta(t)-b_a(t)-b_\lambda(t)\right)\drow N(0,1)\nonumber.
\end{equation}
Further, from the discussions before Theorem 2.1, we know that $b_\lambda(t)$ and $b_a(t)=o_p(1)$ are negligible.
Thus, we have
\begin{equation}\label{subnormal1}
	\left\{\B^\textsf{T}(t)\H_N^{-1} \V_{\pi} \H_N^{-1} \B(t) \right\}^{-1/2}\sqrt{\frac{n}{M}} \left(\hat{\beta}_S(t)-\beta(t)\right)\drow N(0,1)\nonumber.
\end{equation}
This finishes the proof of Theorem \ref{subest}  \\


\noindent\textbf{\emph{The proof of Theorem \ref{AoptTh} }} According to the definition of $\H_N^{-1}$ and $\V_{\pi}$,
\begin{align*}
	tr(\H_N^{-1} \V_{\pi} \H_N^{-1}) &\propto tr\Big[ \H_N^{-1} \sum_{i=1}^N  \Big( \frac{1}{\sqrt{\pi_i}}
	\sum_{k=1}^K (I\{ \varepsilon_i < b_{0k} \} - \tau_k)\tilde{\U}_{ik}\Big)^{\otimes 2} \H_N^{-1} \Big]\\
	&= \sum_{i=1}^N tr\Big[ \H_N^{-1} \Big( \frac{1}{\sqrt{\pi_i}}
	\sum_{k=1}^K (I\{ \varepsilon_i < b_{0k} \} - \tau_k)\tilde{\U}_{ik}\Big)^{\otimes 2} \H_N^{-1} \Big]\\
	&= \sum_{i=1}^N \Big\| \H_N^{-1} \Big( \frac{1}{\sqrt{\pi_i}}
	\sum_{k=1}^K (I\{ \varepsilon_i < b_{0k} \} - \tau_k)\tilde{\U}_{ik}\Big) \Big\|^2 \\
	&= \sum_{i=1}^N \frac{1}{\pi_i} \Big\| \sum_{k=1}^K (I\{ \varepsilon_i < b_{0k} \} - \tau_k) \H_N^{-1} \tilde{\U}_{ik} \Big\|^2 \times \sum_{i=1}^N \pi_i\\
	& \geq \Big( \sum_{i=1}^N \Big\| \sum_{k=1}^K (I\{ \varepsilon_i < b_{0k} \} - \tau_k) \H_N^{-1} \tilde{\U}_{ik} \Big\| \Big)^2,
\end{align*}
where the last inequality follows from Cauchy-Schwarz inequality, if and only if $   \Big\| \sum_{k=1}^K (I\{ \varepsilon_i < b_{0k} \} - \tau_k) \H_N^{-1} \tilde{\U}_{ik} \Big\|$ is proportional to $\pi_i$, the equality holds, i.e.
\begin{align*}
	\pi_i=\frac{\Big\| \sum_{k=1}^K ( \tau_k - I\{ \varepsilon_i < b_{0k} \}) \H_N^{-1} \tilde{\U}_{ik} \Big\|}{
		\sum_{j=1}^N \Big\| \sum_{k=1}^K ( \tau_k - I\{ \varepsilon_j < b_{0k} \} ) \H_N^{-1} \tilde{\U}_{jk} \Big\|}.
\end{align*}
This finishes the proof of Theorem \ref{AoptTh}.\\

\noindent\textbf{\emph{The proof of Theorem \ref{LoptTh}}}
\begin{align*}
	tr(  \V_{\pi}  )  &= \sum_{i=1}^N \Big\|     \frac{1}{\sqrt{\pi_i}}
	\sum_{k=1}^K (I\{ \varepsilon_i < b_{0k} \} - \tau_k)\tilde{\U}_{ik} \Big\|^2 \\
	&= \sum_{i=1}^N \frac{1}{\pi_i} \Big\| \sum_{k=1}^K (I\{ \varepsilon_i < b_{0k} \} - \tau_k) \tilde{\U}_{ik} \Big\|^2 \cdot \sum_{i=1}^N \pi_i\\
	& \geq \Big( \sum_{i=1}^N \Big\| \sum_{k=1}^K (I\{ \varepsilon_i < b_{0k} \} - \tau_k)   \tilde{\U}_{ik} \Big\| \Big)^2,
\end{align*}
where the last inequality follows from Cauchy-Schwarz inequality, if and only if $   \Big\| \sum_{k=1}^K (I\{ \varepsilon_i < b_{0k} \} - \tau_k)  \tilde{\U}_{ik} \Big\|$ is proportional to $\pi_i$, the equality holds, i.e.
\begin{align*}
	\pi_i=\frac{\Big\| \sum_{k=1}^K ( \tau_k - I\{ \varepsilon_i < b_{0k} \})  \tilde{\U}_{ik} \Big\|}{
		\sum_{j=1}^N \Big\| \sum_{k=1}^K ( \tau_k - I\{ \varepsilon_j < b_{0k} \} )   \tilde{\U}_{jk} \Big\|}.
\end{align*}
This finishes the proof of Theorem \ref{LoptTh}.\\


\newpage

\end{document}